\documentstyle[prl,aps]{revtex}

\newcommand{\nn}{\mbox{\boldmath$n$}}

\newcommand{\SS}{\mbox{\boldmath$S$}}
\newcommand{\uu}{\mbox{\boldmath$u$}}
\newcommand{\ww}{\mbox{\boldmath$w$}}

\newcommand{\bkappa}{\mbox{\boldmath$\kappa$}}

\newcommand{\bR}{\mbox{$\cal R$}}

\newcommand{\bONE}{\mbox{\boldmath$1$}}

\newcommand{\la}{\langle\!\langle}
\newcommand{\ra}{\rangle\!\rangle}
\draft\begin{document}\title{Generating moment equations in the Doi 
model of liquid--crystalline polymers}\author{P.\ Ilg, I.\  V.\ Karlin 
\thanks{Permanent address: Institute of Computational Modeling, 
Russian Academy of Sciences, Krasnoyarsk, Russia}, 
H.\ C.\ \"{O}ttinger\\ 
{\it ETH Z\"{u}rich, Department of Materials, Institute of Polymers and 
Swiss F.\ I.\ T.\ Rheocenter, CH--8092 Z\"{u}rich, Switzerland}}
\maketitle 
\begin{abstract} 
We present a self--consistent method for deriving moment equations 
for kinetic models of polymer dynamics. The Doi model \cite{D} of 
liquid--crystalline polymers with the Onsager excluded--volume 
potential is considered as an example. 
To lowest order, this method amounts to a simple effective potential  
different from the Maier--Saupe form. 
Analytical results are presented which indicate that this effective 
potential provides a better approximation to the Onsager potential 
than the Maier--Saupe potential.
Corrections to the effective potential are obtained.
\end{abstract}
\section{Introduction}
Kinetic theory is a powerful analytical tool for describing the dynamics 
of dilute and semi--dilute solutions of polymers in terms of a 
diffusion equation for the particle distribution function or, equivalently, 
by the full system of moment equations. 
In general, the moment system has to be truncated at some level.
The problem of the ``closure approximation'' is very 
well--known in the literature, and an enormous amount of suggestions
have been analyzed in the case where each moment couples only 
to a few higher--order moments. 
However, for some kinetic equations, the time evolution of each 
moment couples to an infinite set of higher moments, so that further 
analytical work is often precluded since closure approximations 
are less studied for this case.
In this paper we present a simple method that allows to derive 
moment equations with a finite coupling valid for a wide class of 
kinetic equations.

In order to be specific, we consider a particularly important example: 
the Doi theory of liquid--crystalline polymers (LCP), subject to the 
Onsager excluded--volume potential \cite{D}.
As it is well--known, in this model each moment equation depends on 
an infinite set of higher--order moments.
In the original work \cite{D}, this problem was treated in two steps: 
First, the Onsager 
potential was replaced by a different, phenomenological potential of the 
Maier--Saupe type \cite{MS}, which gives raise to a coupling to the 
next higher moment only. In the second step, the ``decoupling'' 
approximation was used to solve the resulting closure problem 
for the second moment. 
Subsequent extensive studies were focused on improvements of 
the second step \cite{closure1,closure2,leal,closure3}. 
At the same time, we are not aware of improvements on the first step 
and closure approximations are limited to the Maier--Saupe potential 
up to now. 
However, it would be desirable to deal with the true Onsager 
potential, not only because it becomes exact in the limit of low 
concentrations of perfectly rigid rod--like molecules, but also because 
it contains no phenomenological parameters and therefore gives more 
quantitative predictions. 
In addition, the Onsager potential is preferred in the study of the 
influence of flow on the isotropic--nematic transition, since it gives 
a clear--cut prediction of the range of coexistence of the equilibrium 
isotropic and nematic phase, both in stationary \cite{thirumalai} 
and non--stationary \cite{larson} flows.
The method, which we propose in this work, leads to an approximation 
of the Onsager potential, which, to the lowest order, is at the same time 
as simple as the Maier--Saupe potential but also closer to the 
true Onsager potential. 
Moreover, corrections to this approximation can be obtained in 
a systematic manner. 

\section{The Doi model} 
Let $\psi(\uu; t)$ be the probability distribution function for a rigid 
rod--like polymer molecule to be oriented parallel to the unit vector 
$\uu$. 
The time evolution of $\psi$ in the presence of flow and the Onsager 
excluded--volume potential was given by Doi \cite{D} 
and may be written as:
\begin{equation}
\label{D}
\partial_t\psi = -\bR\cdot\left[ \uu\times\left( \bkappa\cdot\uu\psi\right) 
\right] + \bR\cdot\hat D_{\rm r}\psi\bR 
\left(\frac{\delta A}{\delta \psi(\uu)} \right).
\end{equation}
Here $\bR=\uu\times\partial/\partial\uu$ is the rotational operator, 
$\partial/\partial\uu$ the gradient on the unit sphere, $\bkappa$ 
the gradient of the velocity, $\hat D_{\rm r}$ the rotational diffusivity, 
$\delta/\delta\psi$ the functional derivative and $A=A_0+A_1$ is the 
free energy functional per molecule divided by $k_{\rm B}T$,
\begin{mathletters}
\label{A}
\begin{eqnarray}
\label{A0}
A_0&=&\ln \nu -1 + \langle\ln\psi(\uu)\rangle\\
\label{A1}
A_1&=&\frac{U}{2} \la \sqrt{1-(\uu\cdot\ww)^2} \ra.
\end{eqnarray}
\end{mathletters}
$U=2bL^2\nu$ is the reduced excluded--volume, $2b$ and $L$ 
are the diameter and the length of the rod--like polymeric molecules, 
respectively, and $\nu$ is the number of molecules per unit volume. 
Here and below we use the following notations for averages: $\langle 
f(\uu)\rangle=\int f(\uu)\psi(\uu)d\uu$, and $\la f(\uu,\ww)\ra=\int\!\int 
f(\uu,\ww)\psi(\uu)\psi(\ww)d\uu d\ww$. 
$A_0$ describes the loss of entropy with molecular alignment, 
while $A_1$ expresses the Onsager free energy of steric 
interaction in the second virial approximation \cite{ons}.
Following Doi and Edwards \cite{DE}, the rotational diffusivity 
is approximated by
\begin{equation}
\label{DC}
\hat D_{\rm r} \approx \overline D_{\rm  r} = 
D_{\rm r}\left[\frac{4}{\pi}\la\sqrt{1-(\uu\cdot\ww)^2}\ra\right]^{-2},
\end{equation}
where $D_{\rm r}$, the rotational diffusion coefficient for a rod 
in an isotropic, semi--dilute solution of like rods, is related to the 
rotational diffusion constant for a dilute solution, $D_{\rm r0}$, by 
$D_{\rm r}=cD_{\rm r0}(\nu L)^{-2}$ with an empirical coefficient 
$c$. 
Nonlinearity of Eq.\ (\ref{D}) in $\psi$ brought about by the potential 
(\ref{A1}) reflects the mean--field nature of the Onsager theory of the 
excluded--volume effect. 
The self--consistent potential, identified by Doi \cite{DE}, is related 
to the free energy of interaction, 
$V(\uu)=k_{\rm B}T \delta A_1/\delta \psi(\uu)$.
Various phases of the LCP are conveniently described by the order 
parameter $\SS=\langle \uu\uu-(1/3)\bONE\rangle$, where $\bONE$ 
is the unit tensor. 
It is reasonable therefore to look for approximate formulations of the 
dynamics in terms of the order parameter alone. 
However, as mentioned above, the time evolution equation for 
$\SS$ couples to an infinite number of moments of $\psi$.
In the derivation given by Doi, this difficulty was circumvented by 
replacing the Onsager potential (\ref{A1}) by a different,
phenomenological expression of the Maier--Saupe type \cite{MS}:
\begin{equation}
\label{MS}
A_1^{{\rm MS}}=a_0-\frac{a_1}{2}U\SS:\SS,
\end{equation}
where $a_0$ and $a_1$ are parameters independent of $\psi$. 
A further separate treatment of the diffusivity (\ref{DC}) is also 
necessary. A compact presentation of the entire development is given 
by Doi and Edwards \cite{DE}. The Doi model with the Maier--Saupe 
potential (\ref{MS}) constitutes the basic kinetic model of LCP used by 
many authors for analytical studies to derive equations for the order 
parameter. 
As is well--known, the kinetic equation (\ref{D}) with the 
potential (\ref{MS}) does not give a closed equation 
for the order parameter but contains also the higher--order moment
$\langle \uu\uu\uu\uu \rangle$, and therefore constitutes a further 
problem of closure. 
The original Doi approach was based on 
the decoupling approximation for the fourth--order moments of $\psi$ 
in terms of $\SS$. Improvements on the decoupling approximation 
are currently under active research \cite{leal}. 

\section{Generating moment equations}
In this communication, we demonstrate that a different self--consistent 
treatment of the kinetic equation (\ref{D}) is possible. Modifications 
concern only the relaxational part of the Eq.\ (\ref{D}),  specifically, 
the  excluded--volume potential (\ref{A1}) and the diffusivity (\ref{DC}), 
and therefore we consider the case $\bkappa=0$ in the sequel to 
simplify notations. 
Specifically, we employ the cumulant expansion of the potential 
(\ref{A1}) and the diffusivity (\ref{DC}). 
The leading term of this expansion results in an effective potential 
that differs from the Maier--Saupe potential (\ref{MS}), and which 
contains a non-polynomial dependence on the order parameter $\SS$. 

In the second virial approximation, the free energy of interaction, 
$A_1$ can be written as $A_1=(\nu/2)\la \beta(\uu,\ww)\ra$. 
If only excluded--volume interactions are present, 
the second virial coefficient $\beta$ corresponding to the Onsager 
expression (\ref{A1}) is of the form
$\beta(\uu,\ww)=\beta( (\uu\cdot\ww)^2 )$,
with $\beta(x)=2bL^2\sqrt{1-x}$. 
Specifically, expanding $\beta(x)$ in a Taylor series and interchanging 
summation and averaging in this expansion, we get 
$\la \beta(x) \ra = \sum_{n=0}^{\infty} a_n \la x^n \ra$,
where $a_n$ are numerical coefficients.
Each average $\la x^n \ra$ can be represented in terms of 
cumulants $\la x^k \ra_c$ of order $k\leq n$. 
Resummation of the series leads to
\begin{equation}
\label{expSqrt}
\la \beta(x) \ra = \beta(\la x \ra) + 
\sum_{m=1}^{\infty}\frac{1}{m!}\left(\frac{\la x^2 \ra_c}{2}\right)^m 
\beta^{(2m)}(\la x \ra) + \dots,
\end{equation}
where $\beta^{(2m)}$ is the $2m$--th derivative of $\beta$ and 
ellipses denote terms including third or higher--order cumulants as 
factors. Therefore, the functional $A_1$ (\ref{A1}) can be split as 
$A_1=A_1^{(1)} + B$, 
where $A_1^{(1)}$ is the free energy, corresponding to the total neglect 
of second and higher--order cumulants in each term of the expansion,
\begin{equation}
\label{A1MF}
A_1^{(1)}= \frac{U}{2}\sqrt{1-\langle\uu\uu\rangle:\langle\uu\uu\rangle}.
\end{equation}
In terms of the order parameter $\SS$, $A_1^{(1)}$ may be rewritten 
as $A_1^{(1)}=(U/\sqrt{6})\sqrt{1-(3/2)\SS : \SS}$. 
By the mean value theorem, it is easy to see that $A_1^{(1)}$ gives 
an upper bound to $A_1$, $A_1 \leq A_1^{(1)}$, for the present 
case of excluded--volume interactions. 
The functional $B$ contains the higher--order cumulants. 
While all powers of the second cumulants are displayed in 
Eq.\ (\ref{expSqrt}), in general it is not a priori clear whether it is 
more important to keep powers of the second cumulant or higher 
cumulants. 
However, we generally expect the linear term in the second order 
cumulant to be most important.
The corresponding term $m=1$ in Eq.\ (\ref{expSqrt}), 
$A_1^{(2)}=(1/2)\la x^2 \ra_c \beta''(\la x \ra)$,
gives the first non-vanishing contribution to $B$, 
\begin{equation}
\label{A1cor}
A_1^{(2)}=-\frac{U}{16} \la\left[ (\uu\cdot\ww)^2 - 
\langle\uu\uu\rangle:\langle\uu\uu\rangle \right]^2\ra 
\left( 1-\langle\uu\uu\rangle:\langle\uu\uu\rangle \right)^{-3/2}.
\end{equation}
Keeping only the first $N$ cumulants in the expansion (\ref{expSqrt}), 
the functional (\ref{A1}) is approximated by non--linear functions of 
the first $2N$ moments of $\psi$. 
Inserting this approximation in the time evolution equation (\ref{D}) 
amounts to an approximation of the self--consistent Onsager potential 
$V$ in terms of a polynomial of order $2N$ in $\uu$ but with 
non--linear coefficients. 
In this approximation, the time evolution of the $2n$--th moment 
contains only the first $(2n+2N)$--th moments. 
With this, moment equations can be generated, that approach the 
original equations in a systematic way, thereby containing only 
a finite number of moments at each stage. 

\section{Testing the approximation}
Clearly, the above procedure is most valuable if the first terms, 
$A_1^{(1)}$, etc.\ , already provide a good 
approximation to the full expression $A_1$.
While the general validity of $A_1^{(1)}$ as a good approximation to 
$A_1$ is a rather delicate problem, it should be mentioned that it is so 
at least in two limiting cases. Namely, for the isotropic state, the value 
of $A_1^{(1)}$ differs from $A_1$ for less than 5\%, while in the fully 
ordered state the approximation (\ref{A1MF}) becomes exact. 
Moreover, on the submanifold of distribution functions of the form
\begin{equation}
\label{psialpha}
\psi_{\alpha}(\uu) =\frac{\alpha}{4\pi \sinh \alpha} 
\cosh(\alpha \uu\cdot\nn),
\end{equation}
where $\nn$ is an arbitrary unit vector, and $0\leq \alpha \leq \infty$,
the functional $A_1^{(1)}$ turns out to approximate $A_1$ very well 
for all values of the parameter $\alpha$ between the isotropic state, 
$\alpha=0$, and the fully ordered state, $\alpha=\infty$. 
To show this, we plot in Fig.\ 1 the functions $A_1(\alpha)$, 
$A_1^{(1)}(\alpha)$ and $A_1^{(1)}(\alpha) + A_1^{(2)}(\alpha)$, 
that result upon inserting the ansatz (\ref{psialpha}) into (\ref{A1}), 
(\ref{A1MF}), (\ref{A1cor}), respectively. 
Note, that $A_1(\alpha)$, $A_1^{(1)}(\alpha)$ and 
$A_1^{(2)}(\alpha)$ can be calculated analytically. 
For convenience, we plot the functions against the scalar order 
parameter, defined as $S=\sqrt{(3/2)\SS : \SS}$. 
Including $A_1^{(2)}$ does not only reduce the error of the 
approximate value of $A_1$ in the isotropic state to 1.5\%, but 
improves the accuracy of the approximation over the whole range of 
$S$. For comparison, we included in Fig.\ 1 also the free energy 
$A_1^{\rm MS}$, corresponding to the Maier--Saupe expression 
(\ref{MS}), thereby choosing the undetermined constant so that the 
limit of the fully ordered state is matched correctly. 
Note, however, that in any case
$A_1^{\rm MS}$ decays asymptotically like 
$1/\alpha$, for $\alpha \gg 1$, whereas $A_1$ and $A_1^{(1)}$ 
behave like $1/\sqrt{\alpha}$ in this regime. 
We included in Fig.\ 1 also the derivative of the above functions, 
since they are related to the self--consistent potential $V$. 
Fig.\ 1 shows that also the derivative of $A_1^{(1)}$ provides a 
good approximation to the derivative of $A_1$, with correct limiting 
behavior near the isotropic and fully ordered state. 
Note that including the first correction $A_1^{(2)}$ yields excellent 
agreement to the true Onsager prediction. 
The Maier--Saupe potential captures the main features but, besides 
an undetermined constant, shows the wrong behavior near the fully 
ordered state.
The ansatz (\ref{psialpha}), originally proposed by Onsager \cite{ons}, 
is known to approximate the equilibrium distribution very well. 
Therefore we conclude that $A_1^{(1)}$ represents a good 
approximation to $A_1$, at least on a representative subset 
of distribution functions.

\section{Thermodynamic consistency}
It is worth mentioning that the presentation given so far can easily be 
cast into the recently developed GENERIC formalism of nonequilibrium 
thermodynamics \cite{GEN1,GEN2}. 
In the absence of potential forces, the example of rigid dumbbells, 
which are equivalent to the model of rigid rods, is formulated 
within the GENERIC formalism in Ref.\ \cite{GENdumb}. 
The mean field potentials considered above can be included in a 
straightforward manner, if we recognize that 
$A_1=-S_1$, where  $S_1$ is the entropic contribution per molecule 
to the free energy of interaction divided by $k_{\rm B}$. 
Formulating the original model as well as the approximations within the 
GENERIC formalism guarantees that our treatment is in accordance 
with the principles of nonequilibrium thermodynamics. 
This becomes especially important if the present model is considered 
in nonisothermal situations. 
For example, the structure of the GENERIC formalism requires 
the polymeric contribution to the elastic stress to be
\begin{equation}
\label{stress}
\sigma_{\alpha \beta}^{\rm e} = 3\nu k_{\rm B}T S_{\alpha \beta} - \nu 
k_{\rm B}T \langle (\uu\times\bR \frac{\delta S_1}{\delta 
\psi})_{\alpha}u_{\beta} \rangle. 
\end{equation}
Eq.\ ({\ref{stress}) agrees with the result of Doi \cite{DE}, obtained 
upon varying the free energy functional. 

\section{The lowest order approximation}
In the sequel, we will adopt the lowest order approximation
$A=A^{(1)}=A_0+A_1^{(1)}$,
where $A_1^{(1)}$ is given by Eq.\ (\ref{A1MF}), and $A_0$ is given 
by Eq.\ (\ref{A0}). 
This amounts to neglect of all higher order correlations in 
Eq.\ (\ref{expSqrt}), or, equivalently, setting $B=0$. 
Substituting $A^{(1)}$ instead of $A$ into Eq.\ (\ref{D}), we derive
\begin{equation}
\label{DMF1}
\partial_t\psi=\bR\cdot\hat D_{\rm r} \left[ \bR\psi - 
\psi\bR\left(\frac{U\uu\uu:\langle\uu\uu\rangle}{2\sqrt{1 - 
\langle\uu\uu\rangle:\langle\uu\uu\rangle}}\right)\right].
\end{equation}
It is now possible to identify the self--consistent potential as 
\begin{equation}
\label{UMF}
V^{(1)}(\uu)=\left(\frac{U k_{\rm B}T}{2}\right) 
\frac{1-\uu\uu:\langle\uu\uu\rangle}{\sqrt{1 - 
\langle\uu\uu\rangle:\langle\uu\uu\rangle}},
\end{equation}
which can be compared to the expression obtained from inserting 
the Maier--Saupe free energy (\ref{MS}) into Eq.\ (\ref{D})
\begin{equation}
\label{VMS}
V_{\rm MS}(\uu) = a_2 - a_1U k_{\rm B}T\uu\uu:\langle\uu\uu\rangle,
\end{equation}
where $a_2$ is an arbitrary constant.
The normalized equilibrium solutions to the Eq.\ (\ref{DMF1}) are 
$\psi_{\rm eq}^{(1)}=Z^{-1}\exp[-V^{(1)}/k_{\rm B}T]$. 

The rotational diffusivity (\ref{DC}) is related to the free energy of 
interaction, since
$\overline{D}_{\rm r}=D_{\rm r}\left[\frac{4}{\pi}A_1/(U/2)\right]^{-2}$.
Substituting $A_1 = A_1^{(1)}$ gives
\begin{equation}
\label{DCMFS}
\overline D_{\rm r}^{(1)}=(3\pi^2/32)D_{\rm r}[1-(3/2)\SS:\SS]^{-1}.
\end{equation}
The diffusion coefficient $\overline D_{\rm r}^{(1)}$ (\ref{DCMFS}) is  
positive in the entire physically meaningful  range of the order 
parameter $\SS$. 
Expression (\ref{DCMFS}) should be compared with the Doi 
phenomenological result:
\begin{equation}
\label{DCD}
\overline{D}_{\rm rD}=D_{\rm r}[1-(3/2)\SS:\SS]^{-2}.
\end{equation}
While we have not found an argument which of the two powers, $-1$ 
or $-2$, is more consistent, it should be stressed that our derivation of 
the diffusion coefficient does not need any further assumptions or 
adjustable parameters, while the derivation of Eq.\ (\ref{DCD}) 
\cite{D,DE} requires the matching of $\overline D_{\rm r}$, resp.\ 
$A_1^{\rm MS}$ in both, the isotropic and the fully ordered state.
Due to its relation to $A_1^{(1)}$, the diffusion coefficient 
$\overline D_{\rm r}^{(1)}$ (\ref{DCMFS}) has a correct limit in the 
fully ordered 
state ($D_{\rm r}/\overline {D}_{\rm r}^{(1)}=0$ as soon as 
$\SS:\SS=2/3$ in the ordered state), while the opposite limit of the 
isotropic state ($\overline{D}_{\rm r}=D_{\rm r}$) is matched within 
$8\%$. 
Again, the first correction (\ref{A1cor}) reduces the error in this limit 
to less than $3\%$.

If we adopt (\ref{DC}) and 
approximate $\overline D_{\rm r}$ by $\overline D_{\rm r}^{(1)}$ 
(\ref{DCMFS}), the time evolution of the order parameter $\SS$ can 
be derived from Eq.\ (\ref{DMF1}) by the so--called Prager procedure
\begin{equation}
\label{dtS}
\partial_t \SS = -6\overline D_{\rm r}^{(1)} \SS + 6\overline 
D_{\rm r}^{(1)} \frac{U'}{\sqrt{1-(3/2)\SS : \SS}}(\SS\cdot \langle 
\uu\uu \rangle - \SS : \langle \uu\uu\uu\uu \rangle),
\end{equation}
with $U'=U/\sqrt{6}$.
This expression differs from the result of Doi \cite{D} not only in the 
diffusion coefficient and in the reduced excluded--volume $U$ due to 
the undetermined constant in the Maier--Saupe potential (\ref{VMS}), 
but contains a non--polynomial dependence on the order parameter 
$\SS$, which becomes important in the nematic state.

\section{Conclusion}
We have presented a systematic procedure that allows 
to derive approximate moment equations for the Doi model of LCP, 
which contain only a finite number of higher order moments. 
The first approximation for the Onsager excluded--volume interaction 
results in an effective potential (\ref{UMF}) proportional to $\uu\uu$, 
but different from the Maier--Saupe form (\ref{VMS}) and without 
free parameters. 
We find indications, that (\ref{UMF}) approximates the true Onsager 
potential better than the Maier--Saupe potential. For higher accuracy, 
the first correction seems to be the most important contribution. 
All these approximations are in accordance with nonequilibrium 
thermodynamics. 

Note, that we have not addressed the problem of solving the resulting 
kinetic equations or ``closing'' the moment equations. This work is 
currently under preparation. 
Nevertheless, for comparing Eq.\ (\ref{dtS}) to the corresponding 
equation with the Maier--Saupe potential, we follow Refs.\ \cite{D,DE} 
and consider the decoupling approximation 
$\SS : \langle \uu\uu\uu\uu \rangle = \SS :\langle \uu\uu \rangle 
\langle \uu\uu \rangle$. If the order parameter is assumed to be of the 
form 
$S_{\alpha\beta}=S(t)[n_{\alpha}n_{\beta}-\delta_{\alpha\beta}]$, 
the relaxation equation for the scalar parameter $S$ is found to be 
as follows:
\begin{mathletters}
\label{final}
\begin{eqnarray}
\label{MFrelax}
\partial_tS&=&-6\overline D_{\rm r}^{(1)} 
\frac{\partial A^{(1)}(S,U')}{\partial S},\\
\label{AMFS}
A^{(1)}(S,U') & = & \frac{S^2}{2} - \frac{U'}{9}\sqrt{1-S^2} 
\left(1-\frac{3S}{2}+2S^2\right)-\frac{U'}{6}{\rm arcsin}(S),
\end{eqnarray}
\end{mathletters}
where $U'=U/\sqrt{6}$. Due to a non--polynomial character of the  
function $A^{(1)}$ (\ref{AMFS}), the relaxation equation (\ref{MFrelax}) 
differs formally from the Landau--de Gennes counterpart derived by Doi 
for the Maier--Saupe potential. Expansion of the function (\ref{AMFS}) 
around $S=0$  reproduces the result of Doi for the Maier--Saupe 
potential, subject to a renormalization of the strength of the 
excluded--volume potential, and a difference in the coefficient in front 
of the $S^4$ term. 
Moreover, the relaxation implied by $A^{(1)}$ (\ref{AMFS}) is 
qualitatively similar to the one given by Doi result 
and distinguishes the same three regimes. 
For $U<U_1$, $A^{(1)}$ has only one minimum at $S=0$, so that 
the system finally becomes isotropic.
For $U_1 < U < U_2$, a second local minimum 
occurs. The system either becomes isotropic or nematic depending on 
the initial value of $S$.
Finally, for $U > U_2$, the isotropic state becomes unstable and 
the system always approaches a nematic state.

Due to the undetermined constant $a_1$ in the Maier--Saupe potential 
(\ref{VMS}), the Doi theory predicts the values $U_1$ 
and $U_2$ also in terms of $a_1$. On the contrary, 
the self--consistent potential (\ref{UMF}) contains no free parameters, 
so that $U_1=3^{1/4}\sqrt{8}/(\sqrt{3}-1)\approx 6.22$ and 
$U_2=3\sqrt{6}\approx 7.34$ 
may directly be compared to the values in the Onsager theory
$U_1=8.38$ and $U_2=10.67$. 
However, the values of the order parameter $S_1=1/4$ and 
$S_2=1/2$ at $U_1$ resp. $U_2$ in the Doi theory do not depend 
on $a_1$ and can therefore be compared to the values predicted by 
(\ref{UMF}): $S_1=(\sqrt{3}-1)/2\approx 0.37$ and 
$S_2=1/\sqrt{2}\approx 0.71$. 

In Fig.\ 2, the equilibrium order parameter 
$S_{\rm eq}$ is shown as a function of $\nu/\nu_2$. 
The lower solid line shows the prediction of the Doi theory, whereas 
the upper solid line corresponds to the approximation (\ref{UMF}). 
For the Maier--Saupe potential, $S_{\rm eq}\neq0$ is given by
\begin{equation}
\label{S_sqrt}
S_{\rm eq} = \frac{1}{4} + \frac{3}{4}\sqrt{1 -\frac{8\nu_2}{3\nu}}.
\end{equation}
For the free energy (\ref{AMFS}), 
$S_{\rm eq}$ is given implicitly as the solution to the algebraic equation
\begin{equation}
\label{SeqMF}
\frac{1+S_{\rm eq}-2S_{\rm eq}^2}{\sqrt{1-S_{\rm eq}^2}} = 
\frac{\nu_2}{\nu}.
\end{equation}

For large values of $\nu/\nu_2$, the solution (\ref{SeqMF}) 
approaches the value $S_{\rm eq}=1$ and 
asymptotically behaves like the solution of the Doi theory 
$S_{\rm eq} \sim 1 - \nu_2/\nu$, for large $\nu$.
Note, that the decoupling approximation corrects the asymptotic 
behavior of the Maier--Saupe potential near the fully ordered state.

As is well--known, the detailed form of the interaction potential can 
have significant effect on the behavior of the order parameter in the 
nematic phase \cite{straley}. Specifically, the amount of order at 
the transition is known to be much smaller in the Maier--Saupe 
theory than in the Onsager model. 
For comparison, we included in Fig.\ 2 the values of 
the order parameter obtained from minimizing the true Onsager 
free energy numerically \cite{lasher}, where $\nu_2$ now 
corresponds to the true nematic transition. 
Although the analysis of the phase transitions via the dynamical 
approach is affected by the use of the decoupling approximation, the 
prediction of the self--consistent approach is much closer to the true 
Onsager values than is the Maier--Saupe potential. 

Finally, it should be mentioned that approximations to the Onsager potential 
like Eqs.\ (\ref{A1MF}) and (\ref{A1cor}) can also be used in the case of 
potential flows, following the approach of Thirumalai \cite{thirumalai} 
without additional assumptions.

To summarize, we have developed a direct approach to the Doi model 
with the Onsager potential. We have demonstrated that the resulting 
kinetic equation has much in common with the Doi model with the 
phenomenological Maier--Saupe potential. 
Corrections to the approximation developed here can be found in a 
systematic way from Eqs.\ (\ref{expSqrt}) to (\ref{A1cor}) by taking into
account higher order correlations. 
The approach to derive self--consistent moment equations is applicable 
to other kinetic equations which can be cast into the form (\ref{D}).

\newpage
\section*{Figure captions}
\begin{description}
\item[Fig. 1:]
Free energy of excluded--volume interaction for distribution functions 
(\ref{psialpha}) plotted against the scalar order parameter $S(\alpha)$. 
From top to bottom: approximation $A_1^{(1)}$ (\ref{A1MF}), with 
first correction $A_1^{(1)} + A_1^{(2)}$ (\ref{A1cor}), true Onsager 
expression $A_1$ (\ref{A1}) and the Maier--Saupe free energy 
$A_1^{\rm MS}$ (\ref{MS}), when the limits $S=0$ and $S=1$ 
are matched. 
In the inset, the derivative of the above functions is shown as a function 
of $S(\alpha)$. The order of the curves from top to bottom is the same.

\item[Fig. 2:]
The equilibrium order parameter $S_{\rm eq}$ as a function 
of $\nu/\nu_2$. The figure shows the 
behavior due to the Maier--Saupe potential (\ref{S_sqrt}), lower curve, 
and the solution of (\ref{SeqMF}) corresponding to $A^{(1)}$, upper 
curve, in the decoupling approximation. 
Full circles indicate the order parameter for the true Onsager potential 
in the static case (from \cite{lasher}). 
\end{description}

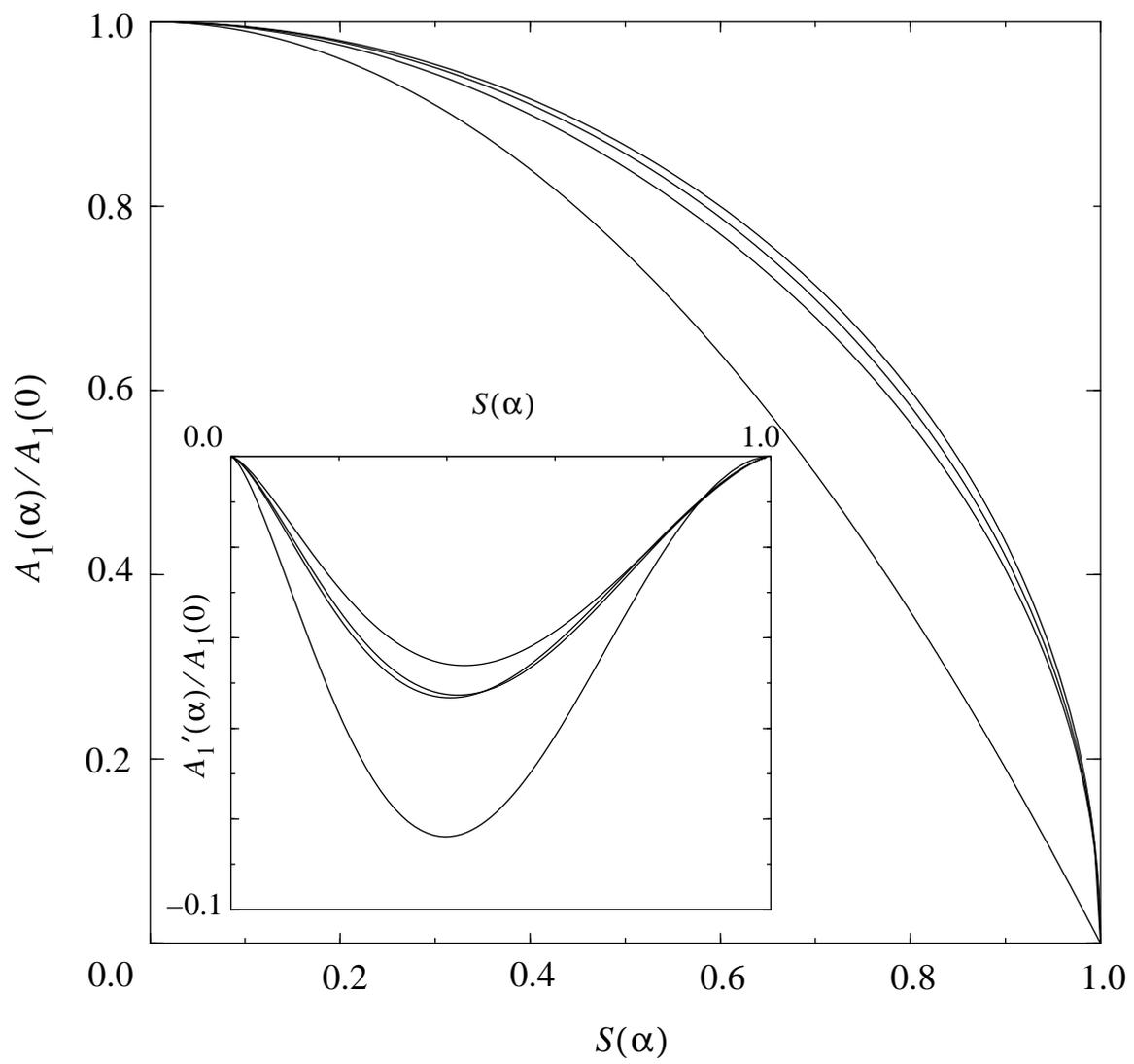
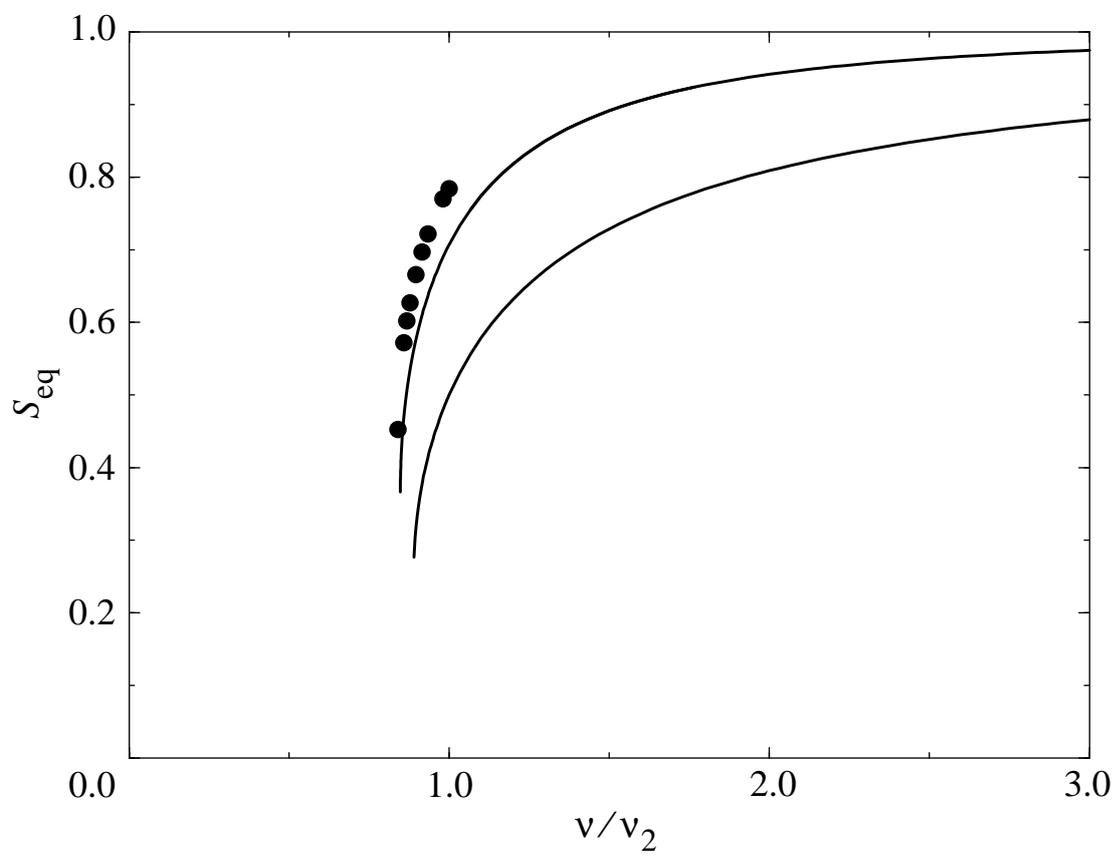
\end{document}